\documentclass[12pt]{article}
\normalbaselines
\catcode `\@=11
\@addtoreset{equation}{section}

\def\section{\@startsection {section}{1}{\z@}{-3.5ex plus -1ex minus
     -.2ex}{2.3ex plus .2ex}{\normalsize\bf}}
\def\subsection{\@startsection{subsection}{2}{\z@}{-3.25ex plus -1ex minus
 -.2ex}{1.5ex plus .2ex}{\normalsize\bf}}

\def\thebibliography#1{\section*{References\markboth
  {REFERENCES}{REFERENCES}}\list
  {[\arabic{enumi}]}{\settowidth\labelwidth{[#1]}\leftmargin\labelwidth
  \advance\leftmargin\labelsep
  \usecounter{enumi}}
  \def\newblock{\hskip .11em plus .33em minus -.07em}
  \sloppy
  \sfcode`\.=1000\relax}
 
\catcode `\@=12
\def\be{\begin{equation}}
\def\ee{\end{equation}}

\newcommand{\sn}{\smallskip\newline}
\newcommand{\mn}{\medskip\newline}

\newcommand{\R}{\mbox{I \hspace{-0.82em} R}}

\newcommand{\p}{{\bf p}}

\begin{document}

\vspace*{2.5cm}
\noindent
{ \bf ON THE ONLY THREE SHORT-DISTANCE STRUCTURES WHICH 
CAN BE DESCRIBED BY LINEAR OPERATORS}\vspace{1.3cm}\\
\noindent
\hspace*{1in}
\begin{minipage}{13cm}
Achim Kempf \vspace{0.3cm}\\
Department of Applied Mathematics and Theoretical Physics\\
and Corpus Christi College, 
 Cambridge University, United Kingdom \\
\end{minipage}

\vspace*{0.5cm}

\begin{abstract}
\noindent
We point out that if spatial 
information is encoded
through linear operators $X_i$, 
or `infinite-dimensional matrices' with an
involution $X_i^*=X_i$ then 
these $X_i$ can only describe either continuous, discrete
or certain "fuzzy" space-time structures. We argue that the
fuzzy space structure 
may be relevant at the Planck scale.
The possibility of this fuzzy space-time structure is 
related to subtle features of infinite dimensional
matrices which do not have an analogue 
in finite dimensions.
For example, there is a slightly weaker version of
 self-adjointness:
 symmetry, and there is 
a slightly weaker version of unitarity: isometry. 
Related to this,
 we also speculate that the presence of horizons 
may lead to merely isometric rather than unitary time evolution.
\end{abstract}
\vskip-12.3truecm
\hskip9.3truecm
{\tt DAMTP-1998-38, hep-th/9806013}
\vskip13.3truecm
\section{\hspace{-4mm}.\hspace{2mm}NONCOMMUTING LIMITS AND
INFINITE-DIMENSIONAL MATRICES}
Since limits do not have to be interchangeable, 
infinite-dimensional matrices possess subtle features
without analogues among finite dimensional matrices.
Let us first look at the trivial example
\be
k(n_1,n_2,a) = \frac{n_1 f(a) + n_2 g(a)}{n_1 +n_2},
\qquad n_1,n_2 \in \mbox{I$\!$N}
\label{hi}
\ee
where $f$ and $g$ are regular functions. Now $ \lim_{n_1\rightarrow\infty}
\lim_{n_2 \rightarrow \infty} k (n_1,n_2,a)= g(a)$. It seems that
when $n_1$ and $n_2$ are taken to infinity
the information on $f$ is lost. Of course it
is only hidden and is not lost, because
 we can find $f$ at infinity by approaching infinity on other paths,
e.g. $ \lim_{n_2\rightarrow\infty}
\lim_{n_1 \rightarrow \infty} k (n_1,n_2,a)= f(a)$. In this sense,
infinity is able to store away things in such
a way that various information can be retrieved by checking
at various corners at infinity.
\sn
Now take an infinite dimensional matrix $X_{ij},  (i,j=1,2,...,\infty)$ which
obeys $X_{ij}=X_{ji}^*$. 
On the set $D$ of all vectors which
have only a finite (but arbitrarily large) number of nonzero entries,
 $X$ is clearly symmetric, i.e. all its 
expectation values are real:
\be 
\langle v\vert X \vert v \rangle
= \sum_{n_1,n_2=1}^{\infty} v^*_{n_1} X_{n_1n_2} v_{n_2}~
 \in \R \qquad ~\forall v\in D
\label{sym}
\ee
To take matrix elements of $X$ between \it generic \rm
vectors is, however, nontrivial: The scalar product of 
$\vert v \rangle$ with 
$\vert X.w\rangle $ does not necessarily coincide with
the scalar product of $\vert X.v\rangle$ with 
$\vert w\rangle$, because for
\be
k(v,w,n_1,n_2) = \sum_{i=1}^{n_1} \sum_{j=1}^{n_2} v^*_i X_{ij} w_j
\label{ll}
\ee
the two limits $n_1 \rightarrow \infty$ and $n_2
 \rightarrow \infty$ have no reason to commute,
i.e. in general we have 
\be
\lim_{n_1\rightarrow \infty} \lim_{n_2 \rightarrow \infty}k(v,w,n_1,n_2) \ne
\lim_{n_2\rightarrow \infty} \lim_{n_1 \rightarrow \infty}k(v,w,n_1,n_2) 
 \label{sa}
\ee
even if both $v$ and $w$ are square 
summable\footnote{
Technically, this is of course the fact that
a symmetric operator $X$ may or may not
have a self-adjoint extension and if it has, it
may have a whole family of self-adjoint extensions.
This will be addressed in detail below.}.
We are now aware that in this case, 
roughly speaking, some information contained in $X$ may be
somewhat nontrivially stored.
\sn
Let us ask whether such subtleties could play a role
in whatever will eventually be the fundamental theory of
quantum gravity.
\section{\hspace{-4mm}.\hspace{2mm} GRAVITY AT THE 
PLANCK SCALE 
 AND NONCOMMUTATIVE GEOMETRY}
Gravity has its fingers
at both ends of the length spectrum:
the extrapolation of general relativity and quantum 
theory, as we know them, 
leads to apparent paradoxes both at small and at large scales.
A truly fundamental theory of quantum gravity will have to 
resolve these problems and explain such tricky
issues like the structure of
space-time at the Planck scale or also the
apparent contradiction of unitary time evolution with
black hole information loss.
\sn
Instead of going into the details of any particular
approach to quantum gravity, such as string theory,
let us step back for a moment to
look from some distance at this problem of 
unifying quantum theory and gravity.
With the above issues in mind, can we see any
assumptions that are conventionally taken for granted in
quantum theory, which might prove to be
relaxed in a fundamental theory of quantum gravity?
\medskip\newline
One suggestion, going back to the 1940s \cite{snyder} is that spatial
information will still be encoded through linear operators,
say $X_i$, but that
these may eventually prove to be noncommutative.
The idea of noncommutative geometry 
has in the meanwhile led to the development of beautiful
and powerful mathematics and has
been supported by new theoretical 
evidence within various approaches to quantum gravity, see e.g.
\cite{e}.
\mn
Let us now consider the possibility that the $X_i$,
could also be non-self-adjoint,
merely symmetric operators\footnote{This 
will also yield a possibility for a more subtle form
of noncommutativity of the $X_i$, which is not seen on the
algebraic level.}.
Correspondingly, the suggestion is that 
certain transformations which one would normally expect
to be unitary, may ultimately prove to be merely 
isometric. 
Let us look at the likely properties of an 
individual $X_i$ more closely:
\section{\hspace{-4mm}.\hspace{2mm} LINEAR $X_i$ CAN ENCODE ONLY
THREE DIFFERENT SHORT DISTANCE STRUCTURES}
We expect that
in whatever algebra the $X_i$ may be found to live, 
the involution in the algebra acts as $X_i^*  =  X_i$.
This is to ensure that all formal
expectation values of the $X_i$ in 
Hilbert space representations are real. One might therefore
be tempted to conclude
that the $X_i$ are self-adjoint
operators in Hilbert space representations. 
Let us, however, recall the following crucial point:
An infinite
 dimensional matrix $X_i$
of which all expectation values are real is called symmetric,
but it is not necessarily self-adjoint, as we just recalled above.
\mn
Generally, a matrix $X$ which has exclusively
real expectation values (i.e. which is by definition
symmetric), describes one out
of exactly three different spatial structures:
\begin{enumerate}
\item{if symmetric and self-adjoint, $X$ may have a 
continuous spectrum and thus describe a 
continuous space, or}
\item{if symmetric and self-adjoint, $X$ may have a 
discrete spectrum, thus describing a lattice, or}
\item{if simple symmetric, $X$ has no eigenvectors, and
describes, as we will see, a "fuzzy" space.}
\end{enumerate}
These are the extreme possibilities and arbitrary mixtures
 of the three can occur: A
 general symmetric matrix can be self-adjoint on subspaces and
a general self-adjoint matrix can
have a mixed continuous and discrete
(even fractal) spectrum.

Of course, this is a statement for generic symmetric operators
which need not be represented as infinite dimensional matrices.
But since we are dealing with separable
Hilbert spaces, let us think
of these operators in matrix representations.
I hope this helps 
intuition by clearly keeping
those infinities that are important here apart from the 
trivial infinitesimals and infinities of continuous representations 
which would be mere artifacts of having
chosen a continuous representation.

\section{\hspace{-4mm}.\hspace{2mm} HOW TO IDENTIFY THE THIRD
SHORT DISTANCE STRUCTURE IN PRACTISE}
Characteristic for the third type short distance structure
is that a simple symmetric $X$ does not have eigenvectors
(if it had eigenvectors, it would be self-adjoint 
on the eigenspaces, but by definition
simple symmetric
operators are not self-adjoint even on subspaces).
In order to identify the third short distance structure in practise,
it would therefore be useful to have a functional which indicates 
the presence or absence of eigenvectors. Indeed, 
a suitable indicator functional is the smallest 
formal uncertainty in $X$, 
defined for any symmetric $X$ on its dense domain
$D_X$ in a Hilbert space as:
\be
(\Delta X)_0 := \inf_{    \{ v \in D_X \vert \langle v \vert v\rangle=1 \} }
\langle v\vert (X- \langle v\vert X \vert v\rangle)^2\vert v\rangle^{1/2}
\ee
The term $\langle v\vert (X- \langle 
v\vert X \vert v\rangle)^2\vert v\rangle^{1/2}$
vanishes for eigenvectors,
and only for eigenvectors. Thus, 
for types 1, 2 and the mixed cases, we have 
$(\Delta X)_0 =0$. Alternatively, when
the formal lower bound to spatial resolution
$(\Delta X)_0$ is some value larger than zero, then we have that
$X$ is of the third type - describing an in this sense
 "fuzzy" geometry.

The stringy correction to the uncertainty relation 
(see e.g. \cite{witten}) provides an example:
\be
\Delta x \Delta p \ge \frac{\hbar}{2}(1 + \beta 
(\Delta p)^2 + ...)
\label{sucr}
\ee
As is easily verified, Eq.\ref{sucr} implies that the $x$-uncertainty is 
finitely bounded from below: $(\Delta X)_0 = \hbar 
\sqrt{\beta}$. Therefore, any operator $X$ that obeys 
this uncertainty relation is of the third type, i.e. it is simple symmetric.
\section{\hspace{-4mm}.\hspace{2mm} INDICATIONS THAT 
FUZZY GEOMETRIES ARE UV REGULAR
AND COMPATIBLE WITH EXTERNAL SYMMETRIES}
The `fuzzy' geometries described
 by $X_i$ which are of the third type 
appear to have attractive features.

As opposed to the second case, the lattice, 
the geometries of the third type 
may arise without breaking external symmetries:
For example the uncertainty relation Eq.\ref{sucr} 
can be obtained from the 
commutation relation 
\be
[X,P]=i\hbar (1+\beta P^2 + ...)
\ee
for which
\be
X \rightarrow X + c, ~ P\rightarrow P
\ee
is an
algebra morphism. This proves that
translation invariance \it can \rm be preserved on third type geometries,
even if, e.g,  Eq.\ref{sucr} should ultimately arise from 
some more sophisticated fundamental theory. 
\sn
Similarly, it has been shown that in more dimensions also rotation invariance
can be preserved: One can find examples of $X_i$ which obey
translation and rotation invariant commutation relations that lead to
uncertainty relations which imply that the $X_i$ are of the third type
(also, if the simple symmetric $X_i$ commute,
the form of the lowest order correction terms to the CCRs is
unique with the fundamental length scale $(\Delta X)_0$
as the only free parameter), see \cite{ak-osc}.
\sn
As opposed to the first case, the continuum, the fuzzy geometries 
appear to also provide a natural ultraviolet cutoff: 
\sn
The framework of quantum field theory may well only have limited validity,
but it is at least an interesting indication that QFTs 
can be  formulated on fuzzy geometries and are ultraviolet
regular, see e.g. \cite{ak-gm-prd}: Technically, when Feynman rules
are defined in position space the ultraviolet divergencies
appear as the ill-definedness of products of propagators and vertices. \sn
These products are
 ill-defined because they are then products of distributions  
like 
\be
\delta(x-x^\prime)=\langle x\vert x^\prime\rangle~ \mbox{ and }~
G(x,x^\prime)=\langle x\vert 1/(\p^2 +m^2)\vert x^\prime\rangle .
\ee
But why are
 $\delta$ and $G$ distributions? They are distributions exactly because 
 the vectors of maximal spatial localisation
`$\vert x\rangle$'
are 
non-normalisable for $X_i$ which describe a continuum, i.e. which are
of the type 1. To be precise:
a vector of maximal 
localisation obeys  $(\Delta X_i)_{\vert 
x\rangle} =(\Delta X)_0, \forall i$ 
and $x$ denotes the formal expectation 
value $\langle x\vert X_i\vert x\rangle
=x_i$. On the continuum they are of course 
the non-normalisable formal $X_i$-eigenfields.
\sn
Clearly, 
the vectors of maximal localisation are normalisable in the case that the
$X_i$ is of the second type, the lattice. 
But, it has also been shown that they are  
normalisable when the $X_i$ are of the third type \cite{go}. If the 
$\vert x\rangle$ are normalisable, $\delta$ and $G$ are regular functions
and their products are well-defined, from which follows that the
 UV divergencies are regularised.  
\sn
While this shows that geometries described by
$X_i$ of the third type  
can preserve external symmetries and regularise the ultraviolet, we still
have not picked up on our observation in the beginning
that with these simple symmetric $X_i$, `some information may be
stored somewhat nontrivially'. 
\section{\hspace{-4mm}.\hspace{2mm} THE DEGREES OF FREEDOM 
CUT OFF BEYOND THE PLANCK SCALE CAN REAPPEAR 
AS INTERNAL DEGREES OF FREEDOM}
If a fundamental theory of quantum gravity does encode spatial
information through $X_i$ of the third type
then those spatial degrees of freedom
that are being cut off beyond the scale $(\Delta X)_0$ may reappear 
in a different form, as internal degrees of freedom:
\mn
Let us first look at QED. It can be formulated by 
specifying that the fields in the, say euclidean, field theoretical
 path integral are in a representation of the algebra
$[P_i, P_j]=F_{ij}(X), ~[X_i,P_j]=i\hbar\delta_{ij}, ~
[X_i,X_j]=0$. The gauge transformations are the group $G$ of
unitaries in the algebra of the $X_i$. 
We note that each unitary, say  $e^{i f(X)}$,
can be written as a function of $n$ elementary unitaries defined as
\be
S_j=(X_j-i1)(X_j +i1)^{-1}
\ee
i.e. there always exists a $g$ such that
\be
e^{i f (X)} = g(S_1, ..., S_n)
\ee 
To obtain a non-abelian 
gauge theory the isospinor structure would normally have to be
introduced `by hand'.
\mn
Now if a fundamental quantum gravity theory does encode spatial 
information through fuzzy $X_i$ then 
the group $G$ of unitaries in the algebra of the $X_i$
is generically non-abelian, even
if  the $X_i$ are commutative:
\sn
The reason is that the $S_i$ which arise from the $X_i$ are merely
isometric and require unitary extension. Generically, unitary
extensions exist and differ by elements of some unitary group $T$
(the unitary automorphisms of the 
deficiency spaces 
 $L^+_i = D_{S_i}^\perp, ~ L^-_i= R_{S_i}^\perp$
of the $X_i$), see \cite{nlis}. The resulting $G$
therefore naturally splits into conventional
unitaries which are functions of the $S_i$, and $T$. As will be shown
 in general and in detail in \cite{aknext}, at
 low energies the gauge transformations defined as the set of unitaries  
which commute with the $X_i$ factors into these internal and external
degrees of freedom, while the 
full $G$ is indeed the full unitary group - the difference between
those internal and external degrees of freedom disappears
at high energies.
\sn
Technically,
the unitary extensions correspond of course to self-adjoint extensions
of the fuzzy $X_i$ beyond their domain of definition (beyond e.g. where
the stringy uncertainty relation holds).
In terms of infinite dimensional matrices, one thereby 
adds to the domain of an $X$ a maximal set of 
vectors for which the limits
$n_1\rightarrow \infty, n_2\rightarrow\infty$ of Eq.\ref{ll} commute. 
Intuitively, these vectors (because they 
have an infinite number of nonzero entries) 
explore the matrix $X_{ij}$ at $i,j =\infty$.
\sn
Within any one self-adjoint extended domain the limits  $n_1,n_2\rightarrow\infty$
commute i.e., in the
jargon of the first page, the vectors in this
domain all probe $X_{ij}$ at the same corner of the infinity $i,j=\infty$.
The set of different self-adjoint domain 
extensions forms a representation of $T$, and each extension probes the 
behaviour  of  $X_{ij}$ at different corners of the infinity $i,j=\infty$.
\sn
Thus, while the fuzzy $X_i$ show only those degrees
of freedom that correspond to structure larger than the cutoff scale
$(\Delta X)_0$, the unitaries in $G$, being bounded operators,
live on the entire
Hilbert space and they therefore must `see' all degrees of freedom.
If a  fundamental quantum
gravity theory does involve fuzzy $X_i$, then it seems, therefore, that
the degrees of freedom beyond the cutoff scale
naturally reappear as internal degrees of freedom,
with a gauge group structure related to $T$.
\section{\hspace{-4mm}.\hspace{2mm} REMARKS ON THE POSSIBILITY OF
SYMMETRIC OR ISOMETRIC OPERATORS OCCURING IN LARGE SCALE GRAVITY}
Finally, let us end with some speculations on the possibility of 
simple symmetric or isometric matrices appearing with 
quantum gravity effects at large scales.
\mn
We recall that the notion
of a particle's energy-momentum is a delicate one.
The fact that there exists a covariant definition of a phase
space of fields and their conjugate \it momentum fields \rm
does not (since momentum \it fields \rm  
act as translators in the space of field amplitudes
rather than as translators on the space-time manifold)
appear to resolve a fundamental paradox:
\sn
Classically, the momentum of a particle is the \it local \rm 
tangent to its geodesic trajectory, while quantum
theoretically a particle 
(as far as the notion of particle goes on curved space) 
which has a fairly certain momentum is 
highly \it delocalised\rm. On flat space one can add and average the momenta
on different virtual trajectories to obtain momentum expectations.
But on an even only slightly 
curved space this summation must, at least,
 pick up uncertainties -
because the tangent vectors live in different
tangent spaces and parallel transport is then
nonunique since path dependent.
Thus, the more a space is curved, 
the more the notion of a particle's 
energy-momentum is imprecise. 
\sn
One may therefore be tempted to give up on a notion of
a particle's momentum on
a curved space. After all, on flat space one is accustomed to the
existence of plane waves which stand for the "points" that make up
momentum space, while on a generic curved space there is no covariant
notion of plane waves.
\sn
On the other hand, there is the
Einstein equation: while the curvature on the
left hand side is  local,
the energy and momentum on the right-hand side are, quantum
mechanically, intrinsically nonlocal, a feature which
 may in some form 
persist to more fundamental, quantised successors of the Einstein
equation. Maybe, one therefore 
ultimately cannot resolve the paradoxes of momenta
on curved space by simply abandoning the concept of momenta. 
\sn
Let us here only speculate that a notion of `fuzzy' momentum operators
as some form of translators on curved space
may exist. We would expect them to be noncommutative to account for
curvature, and we may expect them to be
simple symmetric to account for the absence of
plane waves, i.e. for the absence of points in momentum space.
Finite translations in space or time could then be unitary or
merely isometric. In particular, in the presence of horizons one
may expect an isometric time evolution which does not have a unitary
extension. An isometric time evolution operator would be able to
preserve the scalar product on all fields where it is defined, without, however,
being invertible. Some information may again be `nontrivially stored', and as
extension problems can always be mapped onto abstract boundary 
condition problems, this information may be viewed as sitting on the horizon.
Technically, we would be dealing with the case of deficiency spaces of
nonequal size. For example, the derivative $-i\partial_x$ 
on the half line is merely symmetric
without self-adjoint extension. It gives rise to obviously 
noninvertible 
finite translations which are however still isometric
i.e. scalar product preserving.
\sn
Properties of symmetric and isometric operators have of course
long been known, see e.g. \cite{glaz}, but I hope to have shown
that their potential for physical applications has not yet been
fully explored.
\newpage\noindent
I am grateful to the anonymous referee for pointing out 
a possible relation between the above picture which involves 
(nonresolvable) lattices  at
the Planck scale, and an old idea by Sakharov. In \cite{sakharov}, 
Sakharov relates gravity
and internal degrees of freedom at the Planck scale in order to motivate 
the Einstein-Hilbert action. His idea is that this action expresses 
a `metrical elasticity'
of space-time that arises from curvature-induced nonperfect matching of 
the zero-point energies in the vacum of the QFT of the nongravitational
(inner) degrees of freedom. See also \cite{misneretal}. The analogy suggested 
by Sakharov  
describes gravitons essentially as the phonons of the complicated vacuum of the internal
degrees of freedom.


\begin{thebibliography}{99}

\bibitem{snyder} H.S. Snyder, Phys. Rev. 71: 38 (1947)

\bibitem{e} A. Connes, \it
Noncommutative Geometry, \rm AP (1994), 
 G. Landi, \it 
An Introduction to Noncommutative Spaces and their Geometry \rm,
Lecture Notes in Physics: Monograph 51, Springer, Heidelberg (1997),
J. Madore, \it An introduction to noncommutative 
differential geometry and its physical applications, \rm  CUP (1995),
S. Majid, \it Foundations of Quantum Group Theory, \rm CUP (1996)
 
\bibitem{c} S. Doplicher, K. Fredenhagen, J.E. Roberts,
Comm. Math. Phys. {\bf 172}, 187 (1995), 
D.V. Ahluwalia, Phys. Lett. {\bf B339}, 301 (1994),
M.-J. Jaeckel, S. Reynaud, Phys. Lett. {\bf A185}, 143 (1994),
A. Connes, M.R. Douglas, A. Schwarz, RU-97-94, hep-th/9711162,
C. Rovelli, Preprint C-97-12-16, gr-qc/9803024 

\bibitem{witten} E. Witten, Phys. Today {\bf 49} (4), 24 (1996),
D.J. Gross, 
P.F. Mende, Nucl. Phys. {\bf B303}, 407 (1988),
D. Amati, M. Ciafaloni, G. Veneziano, Phys.Lett. {\bf B216} 
41, (1989),
K. Konishi, G. Paffuti, P. Provero, Phys. Lett. 
{\bf B234}, 276 (1990), R. Guida, K. Konishi, P. Provero, Mod. Phys. Lett. {\bf
A6}, 1487 (1991),
L.J. Garay, Int. J. Mod. Phys. {\bf A10}, 145 (1995)

\bibitem{ak-osc} A. Kempf, J. Phys. {\bf A30} 2093, (1997)
\bibitem{ak-gm-prd} A. Kempf, G. Mangano, Phys. Rev. {\bf D55},
 7909 (1997)
\bibitem{go} A. Kempf, in Proc. 21ICGTMP Goslar, hep-th/9612082 (1997)
\bibitem{nlis} A. Kempf, Europhys. Lett. {\bf 40} (3), 257 (1997)
\bibitem{aknext} A. Kempf, in preparation
\bibitem{glaz} N.I. Akhiezer, I.M. Glazman, \it Theory of Linear
Operators in Hilbert space\rm, F. Unger Publ., (1963)
\bibitem{sakharov} A.D. Sakharov, Soviet Physics - Doklady, {\bf 12} 
(11), 1040 (1968)
\bibitem{misneretal} C.W. Misner, K.S. Thorne, J.A. Wheeler,
 \it Gravitation\rm, p.285 Freeman \& Co (1973)

\end{thebibliography}
\end{document}